\documentclass[aps, prl,8pt, citeautoscript, superscriptaddress, twocolumn, amsmath]{revtex4-2}
\usepackage{graphicx} 
\usepackage[T1]{fontenc} 
\usepackage{natbib}
\usepackage{siunitx}
\usepackage{bm} 
\usepackage{enumitem}
\usepackage{textcomp}
\setlist{nosep} 
\usepackage[colorlinks=true,citecolor=blue,linkcolor=blue]{hyperref}
\usepackage[utf8]{inputenc} 
\usepackage[dvipsnames]{xcolor} 
\usepackage[normalem]{ulem} 

\setcitestyle{super}

\usepackage{color}

\begin{document}
\title{Lateral Exchange Bias for N\'eel-Vector Control in Atomically Thin Antiferromagnets}
\author{Clément Pellet-Mary}
\email[]{clement.pellet-mary@unibas.ch}
\thanks{These authors contributed equally}
\affiliation{Department of Physics, University of Basel,  Basel, Switzerland }

\author{Debarghya Dutta}
\thanks{These authors contributed equally}
\affiliation{Department of Physics, University of Basel,  Basel, Switzerland }

\author{Märta A. Tschudin}
\thanks{These authors contributed equally}
\affiliation{Department of Physics, University of Basel,  Basel, Switzerland }

\author{Patrick Siegwolf}
\affiliation{Department of Physics, University of Basel,  Basel, Switzerland }

\author{Boris Gross}
\affiliation{Department of Physics, University of Basel,  Basel, Switzerland }

\author{David A. Broadway}
\affiliation{School of Science, RMIT University, Melbourne, VIC 3001, Australia}

\author{Jordan Cox}
\affiliation{Department of Chemistry, Columbia University, New York, NY, USA}

\author{Carolin Schrader}
\affiliation{Department of Physics, University of Basel,  Basel, Switzerland }
\affiliation{Laboratoire Charles Coulomb, Université de Montpellier and CNRS, 34095 Montpellier, France}

\author{Jodok Happacher}
\affiliation{Department of Physics, University of Basel,  Basel, Switzerland }

\author{Daniel G. Chica}
\affiliation{Department of Chemistry, Columbia University, New York, NY, USA}

\author{Cory R. Dean}
\affiliation{Department of Physics, Columbia University, New York, NY, USA}

\author{Xavier Roy}
\affiliation{Department of Chemistry, Columbia University, New York, NY, USA}

\author{Patrick Maletinsky}
\email[]{patrick.maletinksy@unibas.ch}
\affiliation{Department of Physics, University of Basel,  Basel, Switzerland }

\date{\today}

\maketitle


{\bf 
Atomically thin van der Waals (vdW) magnets have emerged as a fascinating platform for the exploration of novel physical phenomena arising from their reduced dimensionality and exceptional material properties. 
Their single-crystalline nature and ultimate miniaturization position them as leading candidates for next-generation spintronic applications. 
Antiferromagnetic (AF) vdW magnets are of particular interest, as they combine the advantages of vdW magnets with the functionality of AF spintronics, offering unique opportunities for ultrafast and robust spintronic devices.
However, the lack of approaches to locally and deterministically manipulate their order parameter -- the N\'eel-vector -- remains a key limitation.
Here, we introduce a fundamentally new paradigm in nanomagnetism, which we term lateral exchange bias (LEB), to achieve N\'eel vector control in bilayers of the vdW AF CrSBr.
We exploit the single-crystalline registry formed by terraced CrSBr samples, where the bilayer N\'eel vector is controlled by LEB from neighboring, odd-layered flakes, whose nonzero magnetization we manipulate using magnetic fields. 
Using this control, we achieve nonvolatile manipulation of magnetic domains and  domain walls in AF CrSBr bilayers, establishing a powerful toolkit for controlling atomically thin AFs at the nanoscale.
Our results challenge conventional views on exchange bias and provide a previously unexplored mechanism for achieving atomic-scale control of AFic order.
Our findings pave the way for the development of advanced spintronic architectures and quantum technologies based on vdW magnets.
}

The order parameter of few-layer vdW magnets can be controlled by a range of external stimuli, including 
magnetic\,\cite{Gong2017a,Huang2017a}
and electric fields\,\cite{Wang2018a,Jiang2018a}, 
doping\,\cite{Jiang2018a,Tabataba-Vakili2024Jun},
pressure\,\cite{Song2019a}, 
and strain\,\cite{Bagani2024a}.
However, current approaches do not provide the sort of local control over order parameters that is required to deterministically create spin textures such as magnetic domain walls, and, in most cases, do not allow bistable switching between nonvolatile states which is a prerequisite for vdW spintronics.
While important first steps have been taken in this direction\,\cite{Chen2024a}, realizing such local control in fully compensated vdW AFs remains challenging, and so far no controlled writing of spin textures has been demonstrated. 



In classical magnetism\,\cite{Meiklejohn1956a, Blachowicz2023a} and thin-film spintronics devices\,\cite{Marti2012Jan,Park2011May}, the exchange bias (EB) phenomenon has been known for decades as an efficient approach to controlling magnetic order parameter, and was instrumental for the development of memory and sensor technologies\,\cite{Nogues2005a}. 
The EB effect is based on interfacial exchange interactions between two adjacent magnetic layers, where the order parameter of a pinning layer, typically an AF, controls the order parameter in the adjacent target layer, typically a ferromagnet (FM).
%
%
Despite intense research on EB in vdW magnets\,\cite{Zhu2020a,Dai2021a,Huang2023a,Chong2024a}, its use for vdW spintronics and order-parameter control in vdW magnets remains challenging due to the weakness of the interlayer EB and difficulties in reproducibly controlling the vdW gap in vertical vdW heterostructures\,\cite{ Puthirath2024}.
%



Here, we introduce a novel concept of EB in vdW AFs that is intrinsically free of these limitations and demonstrate its use for local control of the order parameter and spin-texture initialization in bilayers of the vdW AF CrSBr\,\cite{Lee2021Apr}. 
Our concept, which we term lateral exchange bias (LEB), exploits exchange interactions across the one-dimensional interface between two adjacent regions with different numbers of layers of the same single-crystal vdW magnet for order-parameter control. 
Lateral exchange has previously been observed with FM thin films\,\cite{Luo2019Mar,Liu2023Mar} and vdW heterostructures\,\cite{Dai2022Aug,Huang2023Apr}, but never before to manipulate the AF order.
In our demonstration,  an AF-ordered CrSBr bilayer is adjacent to a trilayer with nonzero magnetization, and magnetic-field control of the trilayer is used to control the Néel vector in the bilayer.


\begin{figure*}
    \centering
    \includegraphics[width=\linewidth]{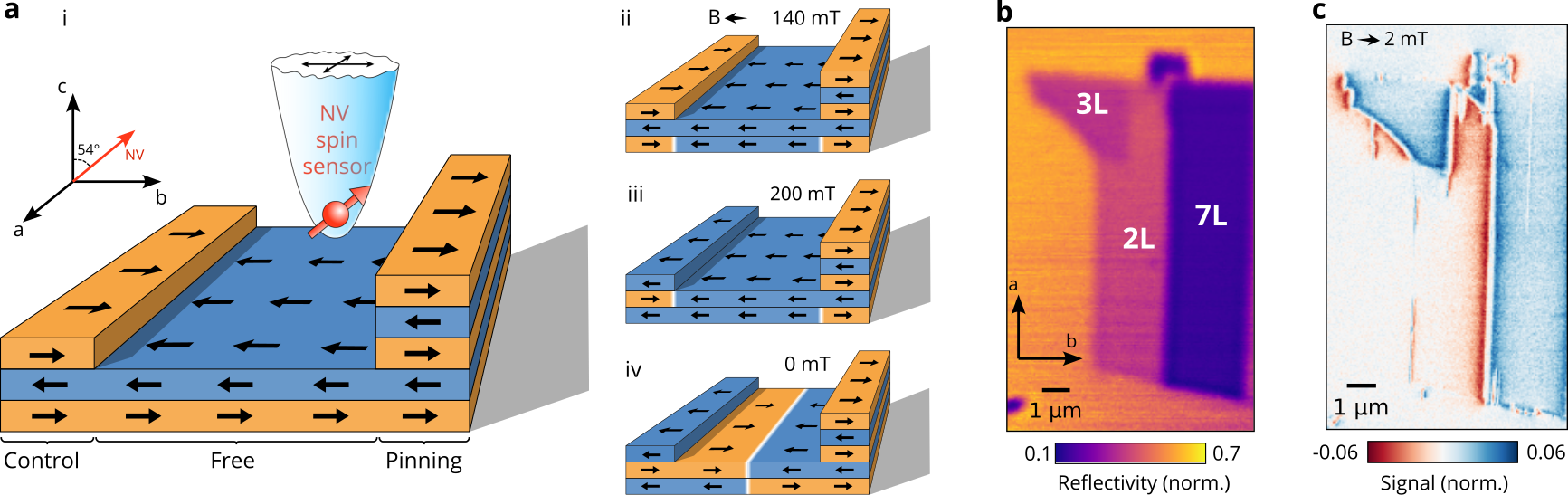}
    \caption{\textbf{Controllable lateral exchange bias.}
    \textbf{a-i} Schematics of the exfoliated CrSBr sample after zero-field cooling. A diamond tip with an embedded NV center, and the angle between the NV center and the sample plane are represented. 
    \textbf{ii-iv} Schematics of the sample after the successive application of 140 mT, 200 mT and 0 mT fields along the crystalline b-axis.
    \textbf{b} Optical image of the flake obtained by measuring the reflectance of a 640 nm laser. Stacks of 2, 3 and 7 layers are labeled. 
    \textbf{c} Dual-Iso-B (see SI section III\,\cite{SI}) magnetic image of the flake for an external in-plane field of $2$~mT. 
    }
    \label{Fig_intro}
\end{figure*}

Figure\,\ref{Fig_intro}\,\textbf{a-i} shows a representative cross-section through our sample that consists of a CrSBr bilayer with adjacent flakes of three and seven layers, respectively (see Fig.\,\ref{Fig_intro}\,\textbf{b} for an optical image), and the expected spin configuration after zero-field cooling of the sample\,\cite{Tschudin2024Jul}. 
CrSBr is an easy axis AF, where the easy axis (the ``$b$-axis'') lies in the vdW plane and where spins order ferromagnetically within a plane and antiferromagnetically between planes\,\cite{Lee2021Apr}.
As a result, CrSBr samples with an even (odd) number of layers carry zero (nonzero) magnetization, respectively.

To assess the magnetic state of our sample, and image its spin textures, we employ direct magnetic imaging using scanning nitrogen-vacancy (NV) magnetometry. 
In short, NV magnetometry exploits the electronic spin of the NV centre as a sensitive magnetometer that can be initialised and read out optically and driven by microwave magnetic fields\,\cite{Rondin2014May}. 
To achieve nanoscale imaging, the NV spin is embedded in a diamond tip\cite{Maletinsky2012May} to be scanned within $\approx50~$nm from the sample. 
In this work, we exploit both qualitative (Dual-Iso-B) and quantitative (ODMR) imaging modes as described in the Supplementary Information (SI) section III\,\cite{SI}, where we use the former for rapid assessment of spin configurations and the latter for detailed quantitative analysis of the resulting spin textures. 
Figure\,\ref{Fig_intro}\,\textbf{c} shows a qualitative NV magnetometry image of our sample that shows stray field patterns consistent with the spin arrangement illustrated in Fig.\,\ref{Fig_intro}\,\textbf{a}-i\,\cite{Tschudin2024Jul}. 
All data we present have been obtained at a temperature of $T \approx 4~$K in a cryogenic scanning NV magnetometry apparatus with vector magnetic field control described elsewhere\,\cite{Thiel2016a, Tschudin2024Jul}. 
When performing NV magnetometry in external magnetic fields, we align these fields with the NV spin quantization axis $\mathbf{e}_{\rm NV}$ to ensure optimal performance\,\cite{Tetienne2012Oct}, and mount our samples such that $\mathbf{e}_{\rm NV}$ lies in the sample's $b$-$c$-plane, tilted $\approx 54~^\circ$ from the $c$-axis (Fig.\,\ref{Fig_intro}\,\textbf{a}-i).
Due to the weak response of CrSBr to magnetic fields along $c$\,\cite{Boix-Constant2022Oct}, we state the magnitude of the $b$-axis component, $B_b$, whenever a field is applied.  

A possible approach to exploit the LEB for N\'eel vector control is shown in the sequence of illustrations in Fig.\,\ref{Fig_intro}\,\textbf{a}, i-iv. 
Following a sequence of increasing magnetic field $B_b$, 
where first, the bilayer undergoes a spin-flip transition at $B_b\approx140~$mT\,\cite{Goser1990Nov,Telford2020Sep,Telford2022Jul,Boix-Constant2022Oct}, and subsequently, the trilayer switches its magnetisation at $B_b\approx200~$mT\,\cite{Tabataba-Vakili2024Jun}. 
Importantly, the $7$-layer flake (that we will refer to as ``pinning layer'' in the following) remains unaffected by this process as the spin-flip field values in few layers CrSBr depend on the exact number of layers, and tend to increase with the flake thickness\,\cite{Ye2022Aug} (see SI section VI\,\cite{SI}).
The flip of the trilayer (that we will refer to as the ``control layer'' in the following) exposes the bilayer to LEB of opposite signs at the boundaries to the control and pinning layer, respectively.
Upon reduction of $B_b$ below $B_{\rm SF}$, AF phases of opposing N\'eel vector orientation will thus emerge from the two boundaries, resulting in the expected final spin arrangement shown in Fig.\,\ref{Fig_intro}\,\textbf{a}-iv. 

\begin{figure*}
    \centering
    \includegraphics[width=\linewidth]{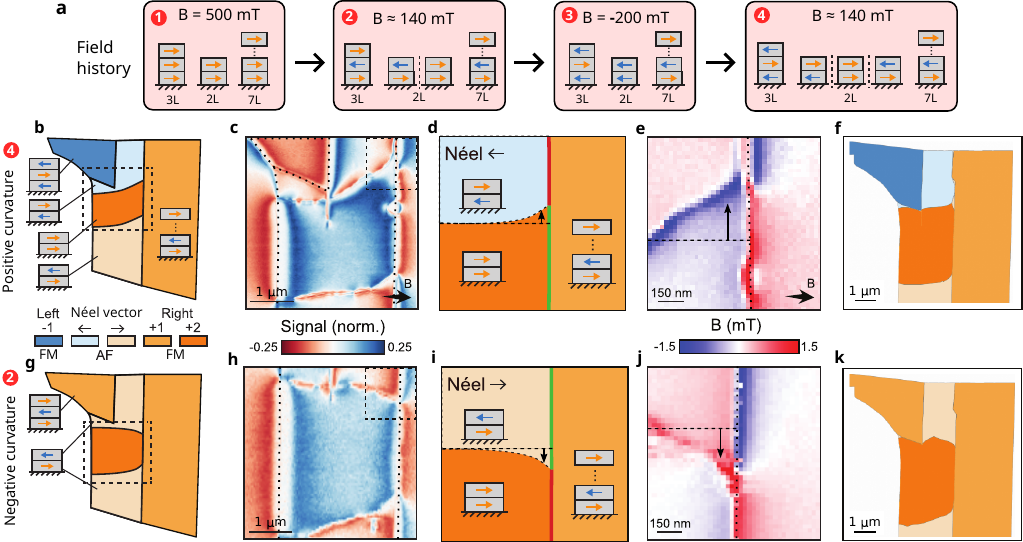}
    \caption{\textbf{Lateral exchange bias induced phase wall} steering. 
    \textbf{a} External magnetic field sequence applied along the easy axis of the flake, with expected magnetization for each stack. \textbf{b-f} correspond to step 4 and \textbf{g-k} to step 2. See SI section VI and VII for more details\,\cite{SI}.
    \textbf{b} Schematics of the magnetic state in the flake for an external field of $144~$mT.
    \textbf{c} Dual-Iso-B magnetic image of the flake in the region highlighted in the schematics. See SI section IV\,\cite{SI} for a full scale imaging of the bilayer. Dotted lines indicate the boundaries between regions of different thicknesses.
    \textbf{d} Illustration of the phase wall distortion at the pinning layer interface. The boundary part painted in red has a higher interfacial energy than the green part. Deviation from normal incidence is noted by a black arrow.
    \textbf{e} ODMR magnetic image of the interface region. Dotted line and black arrow indicate deviation from normal incidence.
    \textbf{f} Micromagnetic simulation of the flake following color code in \textbf{a} (Simulation details in SI section V\,\cite{SI})
    \textbf{g-k} Same as \textbf{b-f} with the control layer's magnetization flipped.}
    \label{Fig_wetting_angle}
\end{figure*}


We begin by experimentally assessing the effectiveness of the LEB. 
For this, we focus on the regime $B_b\approx B_{\rm SF,2}$, where the bilayer is in a state of phase-coexistence between FM and AF phases\,\cite{Tschudin2024Jul,Zur2023Nov}, corresponding to the steps 2 and 4 of the external field sequence presented in Fig.\,\ref{Fig_wetting_angle}\,\textbf{a}. 
Fig.\,\ref{Fig_wetting_angle}\,\textbf{b} shows a schematic of the magnetic state of the flake that we infer from the magnetic image obtained at $B_b =144~$mT (Fig.\,\ref{Fig_wetting_angle}\,\textbf{c}).
Importantly, we find that the LEB has a striking effect on the trajectory of AF-FM phase walls within the bilayer when they impinge on the adjacent pinning layer. 
Specifically, towards the intersection with the bilayer-pinning-layer interface, the phase walls show striking and reproducible distortions away from the normal incidence observed and expected\,\cite{Hubert2008a} when the phase wall impinges on the flake boundary.   
These distortions are well explained by the energetics of the LEB at the bilayer-pinning-layer interface (see Fig.\,\ref{Fig_wetting_angle}\,\textbf{d}):
Where the bilayer has FM spin alignment, the interface incurs an energy penalty corresponding to a head-to-head domain wall\footnote{Note that this description holds true only if the bilayer-pinning-layer interface runs along the crystal $a$-axis, as is the case here.} in a single CrSBr monolayer.
Conversely, if the bilayer is in the AF phase, the interface energy depends on its N\'eel-vector orientation. 
For the arrangement represented in Fig.\,\ref{Fig_wetting_angle}\,\textbf{d}, the interface energy corresponds to a head-to-head domain wall in two CrSBr monolayers, while for the opposite N\'eel vector orientation, the interface energy is zero.
To minimize energy, 
the length of the low-energy bilayer-pinning-layer interface will thus expand at the expense of the high-energy interface. 
This expansion distorts and extends the phase wall, where both represent additional energy penalties.
The process stops when the energy reaches a local minimum, leading to the distorted phase wall trajectory shown in Fig.\,\ref{Fig_wetting_angle}\,\textbf{e}. 
This behaviour can be seen as a magnetic analogy of wetting in hydrostatics, where the phase wall forms a ``contact angle'' with the pinning layer (see SI section VIII\,\cite{SI}).

We confirmed this intuitive picture through numerical micromagnetic simulations described in detail in the SI section V\,\cite{SI}.
Figure\,\ref{Fig_wetting_angle}\,\textbf{e} shows a representative simulation result for the steady state spin configuration in a model CrSBr system that mimics the geometry of our sample. 
Importantly, the simulation yields a phase wall trajectory and FM/AF coexistence that are in good qualitative agreement with our data.

Next, we performed a control experiment in which we investigate the effect of LEB on the phase wall trajectory for the opposite N\'eel-vector orientation compared to the previous case (Fig.\,\ref{Fig_wetting_angle}\,\textbf{g}-\textbf{k}).
We first applied a positive field $B_b=340~$mT to invert the control layer magnetization.
When reducing $B_b$ towards $B_{\rm SF,2}$, an AF pocket preferentially nucleates at the interface to the control layer\,\cite{Tschudin2024Jul}, and, owing to LEB, will have its N\'eel vector flipped compared to the case discussed before.
This assertion is confirmed by the strikingly different phase wall behavior that we observe in this case at $B_b\approx B_{\rm SF,2}$ (Fig.\,\ref{Fig_wetting_angle}\,\textbf{g}). 
The phase wall is now deflected in the opposite deflection compared to before as a result of the modified energetics at the bilayer-pinning-layer interface. 
This picture is again verified by the micromagnetic model as shown in Fig.\,\ref{Fig_wetting_angle}\,\textbf{j}.

\begin{figure*}
    \centering    \includegraphics[width=\linewidth]{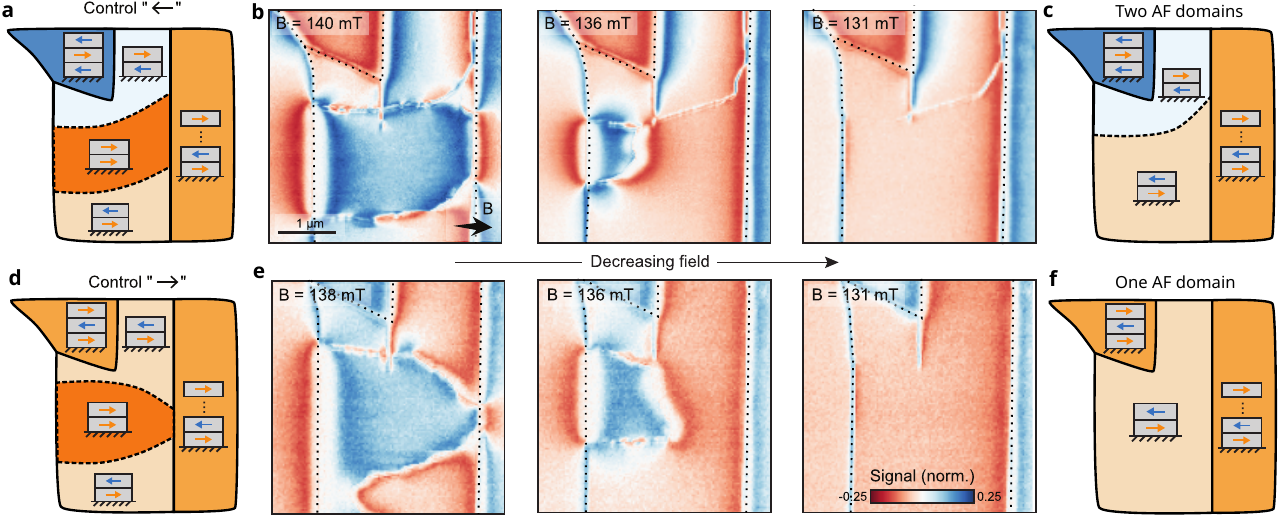}
    \caption{
    \textbf{LEB enabled anti-ferromagnetic domain wall} control. 
    \textbf{a} Schematics of the flake magnetic state at B $\approx$ 140 mT when the control layer is flipped to the left. 
    \textbf{b} Dual-Iso-B magnetic images of the flake as a function of a decreasing external field. Dotted lines indicate the boundaries between regions of different thickness.
    \textbf{c} Schematics of the flake magnetic state at B $\approx$ 130 mT.
    \textbf{d-e} Same as \textbf{a-c} with the control layer flipped to the right.}
    \label{Fig_AFM_domain}
\end{figure*}

We now apply our knowledge of the LEB for the deterministic writing of a domain wall in our AF-ordered CrSBr bilayer. 
Starting from the experimental state shown in Fig.\,\ref{Fig_wetting_angle}\,\textbf{b}, we reduce the magnetic field to below $B_{\rm SF,2}$, where all of the bilayer is AF ordered.
Figure\,\ref{Fig_AFM_domain}\,\textbf{b} shows a corresponding sequence of magnetic images that display the evolution of the bilayer spin texture with decreasing field. 
Strikingly, at $B_b=131~$mT, we find that the bilayer is divided by a well-pronounced line of nonzero stray field. 
Based on our assessment of the bilayer's spin structure we associate this line with the presence of an AF domain wall (Fig.\,\ref{Fig_AFM_domain}\,\textbf{c}). 
A control experiment further confirms this interpretation: 
If the sequence is repeated from the starting configuration presented in Fig.\,\ref{Fig_AFM_domain}\,\textbf{d} no domain-wall would be expected, and indeed, a perfectly homogeneous AF region results in the bilayer (Fig.\,\ref{Fig_AFM_domain}\,\textbf{e}).



\begin{figure}
    \centering
    \includegraphics[width=\linewidth]{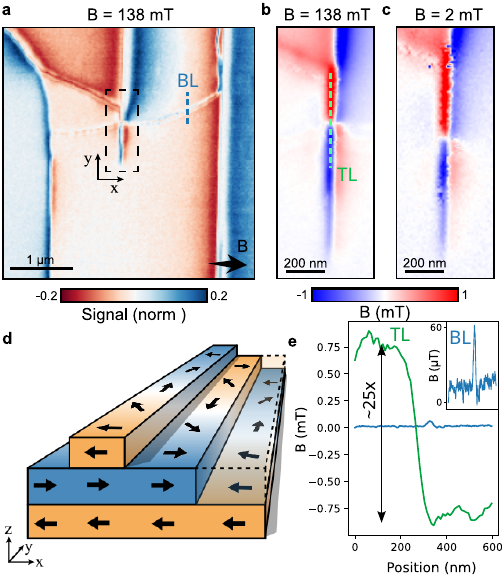}
    \caption{
    \textbf{Decoration of antiferromagnetic order. }
    \textbf{a} Dual-iso-B magnetic image of AF domain wall intersecting a thin FM layer and imprinting onto it.
    \textbf{b, c} ODMR magnetic images of the AF-FM domain wall interface at the initialisation field of $B_b = 138~$mT and with the field reduced to near zero $B_b = 2~$mT, respectively.
    \textbf{d} Schematic of the in-plane spin rotations of each layer at the domain wall interface. 
    \textbf{e} Linecuts of the magnetic field from the bilayer (blue) and trilayer (green) domain walls, as shown in \textbf{a} and \textbf{b}, with an approximate 25$\times$ difference in magnitude. 
    }
    \label{Fig_reverse_EB}
\end{figure}

While a determination of the exact domain wall structure is out of the scope of this work, our simulations indicate that it is of N\'eel nature.
The observed, nonzero domain wall stray field appears to originate from spin-canting induced by the in-plane ($b$-axis) component of the applied field, as its magnitude depends on the domain wall angle (excluding an out of plane canting) and increases with the applied field.
When repeating the experiment, we observed the domain wall at different, random locations on the sample (see SI section IX\,\cite{SI}), which indicates that domain-wall motion is not strongly affected by pinning. 
We also successfully applied this sequence and produced a domain wall on an additional bilayer sample (See SI section X\,\cite{SI}).

Finally, our experiment also revealed an instance of AF domain decoration. Figure\,\ref{Fig_reverse_EB}\,\textbf{a} shows the outcome of a new instance of the domain-wall writing sequence, where the resulting domain wall crosses a $\approx 200~$nm wide CrSBr trilayer that extends from the control layer into the bilayer (see SI section I\,\cite{SI}).
A quantitative magnetometry scan (Fig.\,\ref{Fig_reverse_EB}\,\textbf{b} and \textbf{c}) reveals that, by virtue of the AF interlayer exchange interaction, the domain structure of the bilayer is imprinted onto that trilayer that is now split into a two-domain state (see illustration in Fig.\,\ref{Fig_reverse_EB}\,\textbf{d}).
The resulting twenty fold increase of the domain wall stray field on the narrow trilayer compared to the bulk bilayer (Fig.\,\ref{Fig_reverse_EB}\,\textbf{e}), suggests that such thin stripes act as efficient decorations for underlying AF domains and may thereby enable detection of such domains near zero magnetic field, where the domain wall stray-field is otherwise undetectable.

In conclusion, we have established the LEB as a novel tool for N\'eel-vector control in vdW AFs. 
Importantly, our concept readily extents to other vdW magnets and should, in particular, apply to any $a$-type vdW AF, including the prominent examples CrI$_3$\,\cite{Huang2017a}, CrCl$_3$\,\cite{Wang2019Dec}or CrPS$_4$\,\cite{Son2021Oct}, 
where our ``wetting'' angle methodology could also be applied to assess interfacial exchange energies. 
Our results build on the single crystalline nature of vdW magnets, which offers atomically sharp lateral interfaces — a key factor that sets them apart from their thin-film counterparts and that results in a remarkable enhancement in the reach of EB: 
While in thin-film geometries\,\cite{Marti2012Jan,Park2011May}, each interfacial spin from the pinning layer controls tens of spins in typically nanometres-thin target layers, in our LEB, each interfacial spin controls a row of spins extending microns into the target layer — an extension of the effective range of EB by several orders of magnitude.

Our work opens up exciting future avenues, not only towards the fundamental understanding of domain walls\,\cite{Tetienne2015Apr} and domain formation in atomically thin AFs, but also in the recently explored interplay between magnetism and optical and magnetic excitations of CrSBr\,\cite{Tabataba-Vakili2024Jun}. 
Indeed, excitons and magnons appear strongly coupled in this system\,\cite{Bae2022Sep}, which, together with magnon guiding on spin textures\,\cite{Garcia-Sanchez2015Jun}, might enable engineered magnon-exciton dynamics, controlled by LEB-written AF domain walls. 


\section{Acknowledgment}
During the writing of this manuscript, we became aware of the work by Sun et al.\cite{Sun2025Feb} which uses a complementary technique to ours based on phase coherent second harmonic generation in order to determine the AF order of CrSBr bilayers and tetralayers.

We acknowledge financial support by the ERC consolidator grant project QS2DM, 
by SNF project No. 188521, and from the National Centre of Competence in Research (NCCR) Quantum Science and Technology (QSIT), a competence centre funded by the Swiss National Science Foundation (SNF).
Synthesis of the CrSBr crystals was funded by the Columbia MRSEC on Precision-Assembled Quantum Materials (PAQM) under award number DMR-2011738 and the Air Force Office of Scientific Research under grant FA9550-22-1-0389. Bulk magnetic measurements were supported under Energy Frontier Research Center on Programmable Quantum Materials funded by the US Department of Energy (DOE), Office of Science, Basic Energy Sciences (BES), under award DE-SC0019443. The instrument used to perform these magnetic measurements was purchased with financial support from the National Science Foundation through a supplement to award DMR-1751949.
Micromagnetic calculations were performed at sciCORE (http://scicore.unibas.ch/) scientific computing center at University of Basel.
B.G. acknowledges the support of the Canton Aargau.
DAB acknowledges support from the Australian Research Council through grant DE230100192

\noindent\textbf{Author contributions}

The NV measurements were performed by CPM, DD, MAT together with PS, CS and DAB, under the supervision of PM. 
JH made the NV devices. 
The CrSBr Samples were prepared by JC, and the bulk crystal was synthesised by DGC, all under the supervision of XR and CRD. 
BG provided micromagnetic simulations of the material. 
CPM and PM wrote the manuscript.
All authors discussed the data and commented on the manuscript.

\section{Data availability}
The data that support the findings of this study are available at https://doi.org/10.5281/zenodo.15806866

\section{Code availability}
The codes that support the findings of this study are available from the corresponding authors upon request.

\section{Ethics declarations}
The authors declare no competing interests.

\bibliographystyle{naturemag}
\bibliography{CrSBr_bib} 

\end{document}


\title{SI - Lateral exchange bias}
\author{Clément Pellet-Mary}
\email[]{clement.pellet-mary@unibas.ch}
\thanks{These authors contributed equally}
\affiliation{Department of Physics, University of Basel,  Basel, Switzerland }

\author{Debarghya Dutta}
\thanks{These authors contributed equally}
\affiliation{Department of Physics, University of Basel,  Basel, Switzerland }

\author{Märta A. Tschudin}
\thanks{These authors contributed equally}
\affiliation{Department of Physics, University of Basel,  Basel, Switzerland }

\author{Patrick Siegwolf}
\affiliation{Department of Physics, University of Basel,  Basel, Switzerland }

\author{Boris Gross}
\affiliation{Department of Physics, University of Basel,  Basel, Switzerland }

\author{David A. Broadway}
\affiliation{School of Science, RMIT University, Melbourne, VIC 3001, Australia}

\author{Jordan Cox}
\affiliation{Department of Chemistry, Columbia University, New York, NY, USA}

\author{Carolin Schrader}
\affiliation{Department of Physics, University of Basel,  Basel, Switzerland }
\affiliation{Laboratoire Charles Coulomb, Université de Montpellier and CNRS, 34095 Montpellier, France}

\author{Jodok Happacher}
\affiliation{Department of Physics, University of Basel,  Basel, Switzerland }

\author{Daniel G. Chica}
\affiliation{Department of Chemistry, Columbia University, New York, NY, USA}


\author{Cory R. Dean}
\affiliation{Department of Physics, Columbia University, New York, NY, USA}

\author{Xavier Roy}
\affiliation{Department of Chemistry, Columbia University, New York, NY, USA}

\author{Patrick Maletinsky}
\email[]{patrick.maletinksy@unibas.ch}
\affiliation{Department of Physics, University of Basel,  Basel, Switzerland }
\date{\today}

\maketitle
\tableofcontents
\renewcommand{\thefigure}{S\arabic{figure}}
\setcounter{figure}{0}
\newpage
\section{Sample preparation}

\begin{figure}[h]
    \centering
    \includegraphics[width=0.8\linewidth]{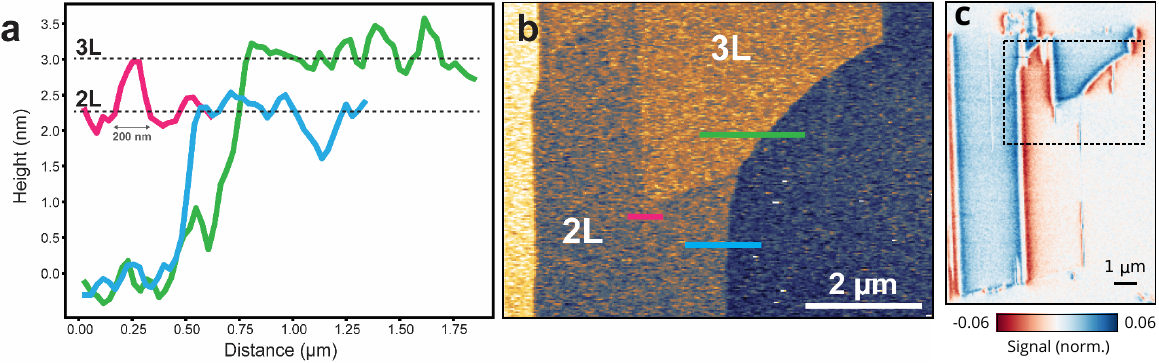}
    \caption{\textbf{a }Atomic force microscopy (AFM) height profiles across 2L (blue), 3L (green) step edges, and small 3L area of interest (red). \textbf{b} AFM image with location of height profiles marked in corresponding colors. \textbf{c} Dual-Iso-B magnetic imaging of the flake showing the region probed by AFM}
    \label{fig:sample AFM}
\end{figure}

\subsection*{Sample preparation methods}
Large single crystals of CrSBr were grown using a chemical vapor transport reaction, as described in Scheie et al.\,\cite{Scheie2022Sep} CrSBr flakes were mechanically exfoliated under ambient conditions onto oxygen plasma cleaned 280 nm SiO2/Si$^+$substrates (NOVA HS39626-WO) with Scotch Magic$^\textrm{TM}$ tape.\,\cite{Novoselov2004Oct,Novoselov2005Jul,Huang2015Nov} Flake thickness was identified by optical contrast and confirmed with atomic force microscopy.\,\cite{Telford2022Jul,Lee2021Apr}

\subsection*{AFM height profile}
Atomic force microscopy was performed in Jupiter XR Asylum Research AFM using AC-240 tips in tapping mode. Height profiles were averaged over 10 pixels and extracted using Gwyddion. 

Fig\,\ref{fig:sample AFM} shows the height profile of the bilayer and trilayer regions mentionned in the main text\, as well as the narrow trilayer mentioned in Fig\,4. This feature was measured to have a width between 100 and 200 nm. Due to differences in tip interactions with the substrate and sample, the step edges between SiO2 and CrSBr are larger ($\approx 1.4~$nm) than the actual heights of the flakes ($\approx 0.8~$nm). Both optical absorption and magnetic analysis confirm the actual number of layers.

\newpage
\section{Experimental setup}

For our experiment, we use the Attocube attoLIQUID 1000 cryostat previously described in \cite{Thiel2019a}. It is a liquid He4 bath cryostat equipped with a 3D-superconducting vector magnet (Janis) with limits of 0.5 T along each (Bx,By,Bz) directions. Broadly, our microscope consists of a scanning AFM based setup in combination with a confocal microscope with optical access at cryogenic temperatures (T=4.2K). 

Two sets of attocube positioners (scanners ANSxyz50 and coarse ANPxyz51) are present on both the tip and sample side. This enables precise and independent movement of the NV-tip and sample. The NV-tip is aligned to the optical axis while the sample is scanned. The NV-tip consists of a diamond cantilever with a parabolic tip having a single NV-center close to its apex. The cantilever is attached to a tuning fork to perform AFM. 

Optical access of the NV-spin is given through a LT-compatible objective (Attocube LT-APO/VISIR/0.82, 0.82 NA) mounted directly on the microscope and a home-built confocal setup. We excite the NV-electronic spin using a 532nm laser (LaserQuantum, GEM532) and the NV-photoluminescence (PL) is collected using a dichroic mirror and measured using an avalanche photodiode (Excelitas, SPCM-ARQH-13). We use a gold wire-antenna, bonded across the sample (distance of ~80 $\mu$m) to send microwave pulses (SRS SG384) for spin manipulation.

In scanning configuration, the standoff distance between the NV-center in the tip and the CrSBr sample is around 50 nm, thereby limiting our spatial resolution to around this value. Furthermore, the presence of a magnetic field that is transverse to the NV-axis leads to mixing of NV-spin states, resulting in decreased ODMR contrast and thereby lower sensitivity. Therefore, we always align our external magnetic field along the NV-axis ($\theta$ = 54$^{\circ}$) resulting in an additional out-of-plane B-component.

\newpage
\section{Magnetic imaging}
\begin{figure}[h]
    \centering
    \includegraphics[width=0.8\linewidth]{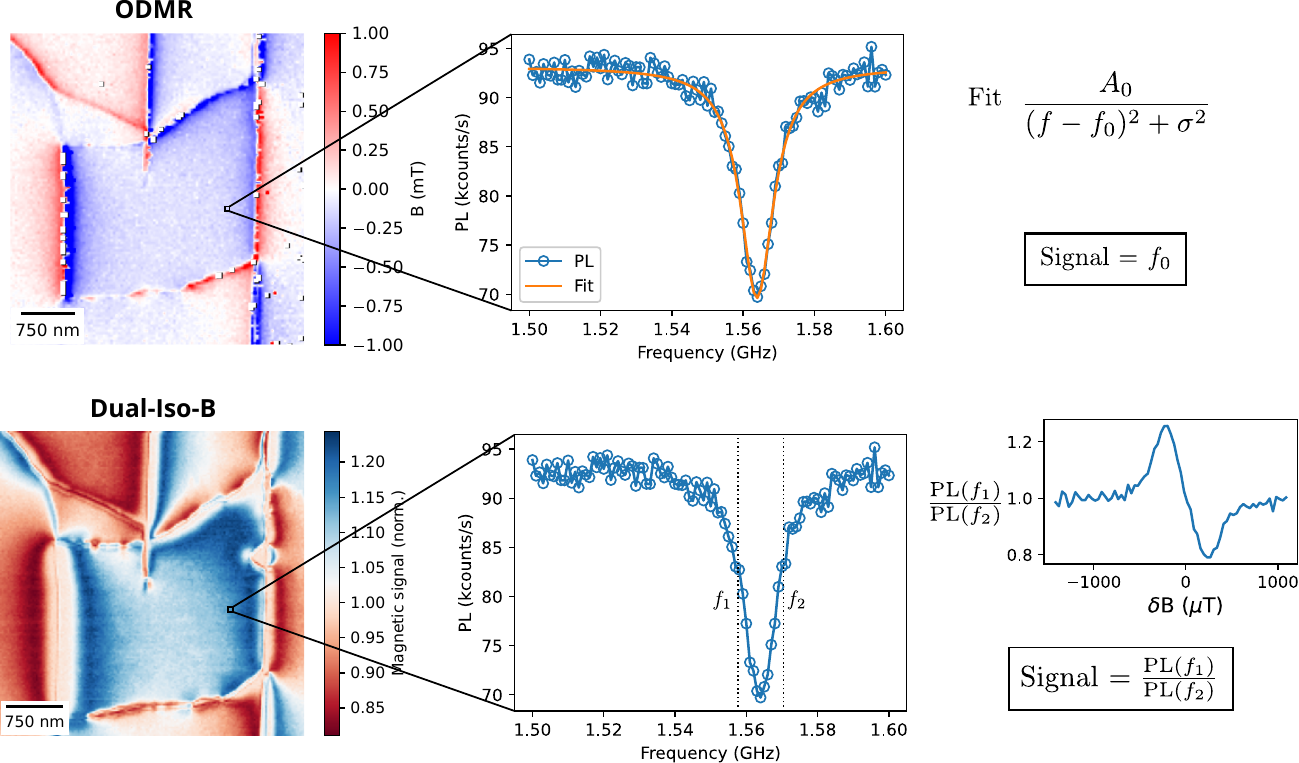}
    \caption{Example of Dual-Iso-B and ODMR magnetic imaging for similar experimental conditions.}
    \label{Fig_Magnetic_imaging}
\end{figure}
Magnetic imaging of our sample is performed using two sensing schemes - Optically Detected Magnetic resonance (ODMR) and Dual Iso-Magnetic field measurement. 

In an ODMR measurement, we continuously excite our NV with a green laser (532nm) and record the photoluminescence (PL) response while driving the spin transition (typically the $m_s = 0$ to $m_s = -1$ transition) using our microwave (MW) source. On resonance, there is a dip in the PL as the NV population is transferred to the darker spin state. The position of this dip is proportional to the magnetic field along the NV-axis. In the vicinity of our sample, the position of this dip changes since the net effective magnetic field is modified, enabling us to extract quantitatively the magnetic contribution from the sample. As we scan over the sample, we measure the ODMR at every point, perform a Lorentzian fit to the spectrum and obtain the additional shift due to the sample. Consequently, we obtain a quantitative measurement of the magnetic stray field from the sample as shown in fig. \ref{Fig_Magnetic_imaging} \textbf{a}. 

Measuring the ODMR spectrum at all points during a scan is time intensive and inefficient as most of the measured points have a weak sensitivity to a change in magnetic field. Dual-Iso-B imaging provides a faster(and therefore more sensitive) measurement scheme. In this technique, an IQ mixer is used to modulate the MW source and generate a lower frequency (f1) and higher frequency (f2) drives, usually with f2-f1 as the width of the ODMR. Independent readout of the two PL values at f1 and f2 and normalizing leads to a single spectrum. Away from the sample, we set the applied frequency such that the measured signal S = PL(1)/PL(2) = 1. In the vicinity of the sample, we observe a shift in S, which can be directly translated to the additional magnetic field from the sample. However, if the stray field from the sample is large, the signal S can be beyond the linear regime of the spectrum. In such a case, this method is no longer quantitative but only provides a qualitative picture based on the direction of the shift of the signal S (Fig. \ref{Fig_Magnetic_imaging} \textbf{b}).

In our experiments, we combine both the quantitative full-ODMR imaging and the qualitative Dual-Iso-B imaging to study the magnetic behavior of CrSBr in different scenarios. In this manuscript, ODMR maps and Dual-IsoB maps are color-coded in the same fashion as shown in Fig.\ref{Fig_Magnetic_imaging}.

To determine the magnetization state from the stray field imaging (either Dual-Iso-B or ODMR), we analyze the stray field sign and magnitude at the edges of the flake and domain walls as explained in previous works \citep{Thiel2019a,Tschudin2024Jul}. Specifically, the stray field profile from the left edge of the bilayer to the substrate yields a magnetisation of zero in the AF region and a magnetisation corresponding to two monolayer magnetisations pointing in the direction of the magnetic field.
Conversely, considering the step between the substrate and the bilayer and the bilayer to the pinning layer, yields the magnetisation strengths and directions of these two layers, respectively. 

\newpage

\section{Large scale imaging}

\begin{figure}[h]
    \centering
    \includegraphics[width=\linewidth]{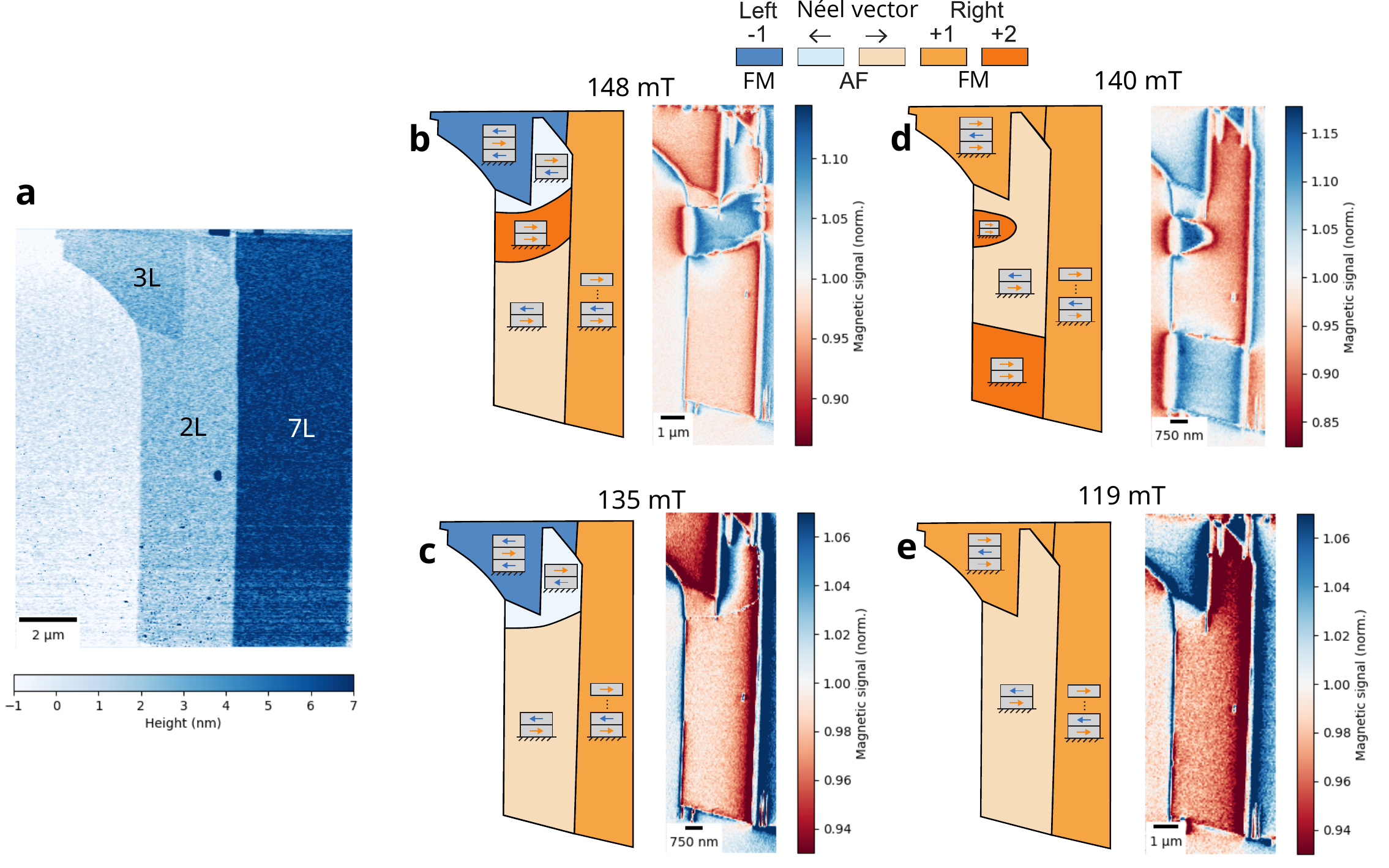}
    \caption{Large scale images of the bilayer and the adjacent control and pinning layers. \textbf{a} AFM height profile of the flake. \textbf{b-e} Dual-Iso-B magnetic images of the bilayer, and schematics of the underlying magnetization, for a configuration with an inverted control layer (\textbf{b,c}) or a parallel control layer (\textbf{d,e}) and different values of the external field.}
    \label{Fig_Full_Flake}
\end{figure}
Most of the magnetic images in this manuscript focus on specific regions of interest. We show here magnetic imaging of the entirety of the bilayer in different cases.

Fig.\ref{Fig_Full_Flake}\,\textbf{a} presents an AFM height profile of the three relevant stacks to this study (bilayer, trilayer and 7-layer), showing no visible structural defects and clean boundaries between the stacks.

Fig.\ref{Fig_Full_Flake}\,\textbf{b} and \textbf{c} show magnetic imaging of the flake for an inverted control layer (in a similar configuration to Fig.3\,\textbf{b} of the main text), for external field values of 148 mT, where part of the bilayer has flipped into a FM stacking configuration, and 135 mT where the bilayer is split into two AF domains.
These images show in particular that no magnetic textures form in the bottom part of the bilayer, as is expected by the fact that this bottom part only shares a boundary with the 7-layer.

Fig.\ref{Fig_Full_Flake}\,\textbf{b} and \textbf{c} show magnetic imaging of the flake for a control layer aligned with the pinning layer (in a similar configuration to Fig.3\,\textbf{d} of the main text). For an external field value of 140 mT, two regions of the bilayer show FM stacking configuration. Both of these domains collapse as the field is lowered to 119 mT, leaving the flake as a uniform, single AF domain.

\newpage
\section{Micromagnetic simulations}
\label{sect:micromagnetics}
\begin{figure}[h]
    \centering
    \includegraphics[width=\linewidth]{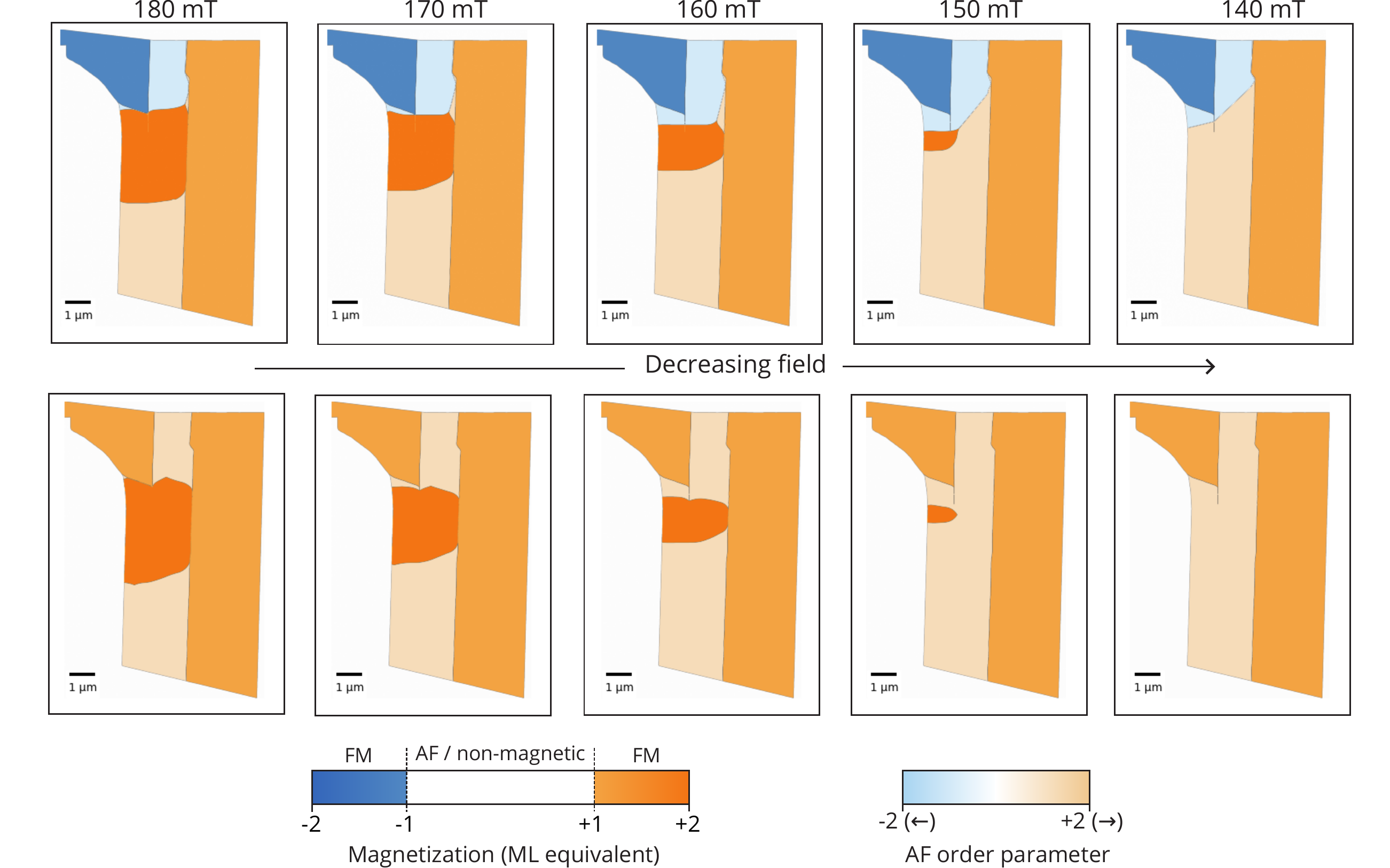}
    \caption{Micromagnetics simulations for decreasing external field. In order to represent both the FM and AF order, a threshold is applied and two distinct color scales are used. The total magnetization ($M_{1L}+M_{2L}+M_{3L}$) is plotted when its absolute value is superior to 0.9, otherwise, the AF order parameter ($M_{1L}-M_{2L}+M_{3L}$) is plotted, where $M_{nL}$ refers to the relative magnetization along the $x$-axis of the $n$-th layer.}
    \label{Fig_micromag}
\end{figure}
The micromagnetic simulations are directly building  on the model presented in Ref.~\onlinecite{Tschudin2024Jul}, employing the same parameters unless otherwise specified. In short, black-and-white images of each layer, derived from optical images of the flake, are utilized to establish the geometry. The layered structure of CrSBr is represented within the finite difference mesh by assigning each cell a thickness equal to that of a single CrSBr layer. Conversely, the inplane mesh cell size is kept on the few-nanometer scale, which allows the simulation of micron scaled flakes. The exchange coupling between layers is set to a small, negative fraction of the in-plane exchange coupling, promoting a-type antiferromagnetic ordering.

The pinning layer (see Fig.1 in the main text), consisting of 7 CrSBr layers, is reduced to 3 layers for simplicity and computational costs, for which we do not expect relevant impact on the investigated physics. The total simulation volume is roughly \SI{8}{\micro\meter} x \SI{12}{\micro\meter} x \SI{2.4}{\nano\meter}, with 1600 x 2400 x 3 cells and corresponding cell sizes of \SI{5}{\nano\meter} x \SI{5}{\nano\meter} x \SI{0.8}{\nano\meter}, respectively. A xy variation in the magnitude of the exchange coupling between layers is engineered to facilitate stable FM domains in the same regions as observed in experiment for similar externally applied magnetic field strengths. We then initialize the magnetic state in the flake in a configuration close to what is experimentally observed for a given applied magnetic field, very similar to what is sketched in Fig.~2 e and j in the main text. The micromagnetic solver~\cite{vansteenkiste_design_2014, exl_labontes_2014} then relaxes the state to the next local energetic minimum that it finds.

Beyond the simulations presented in the main text, we have also traced the evolution of the domain formation with decreasing external magnetic field for both experimental situations sketched in Fig.~2 a and f in the main text. This is shown in Fig.~\ref{Fig_micromag}, which reaffirms the analysis of the evolution presented in the main text. 

We do note that the splitting of the AF domain wall from the AF-FM phase wall observed in the first series is also sometimes observed experimentally (see section \ref{sect:more examples})

\newpage
\section{Spin-flip transitions in few-layers CrSBr}
\begin{figure}[h]
    \centering
    \includegraphics[width=0.8\linewidth]{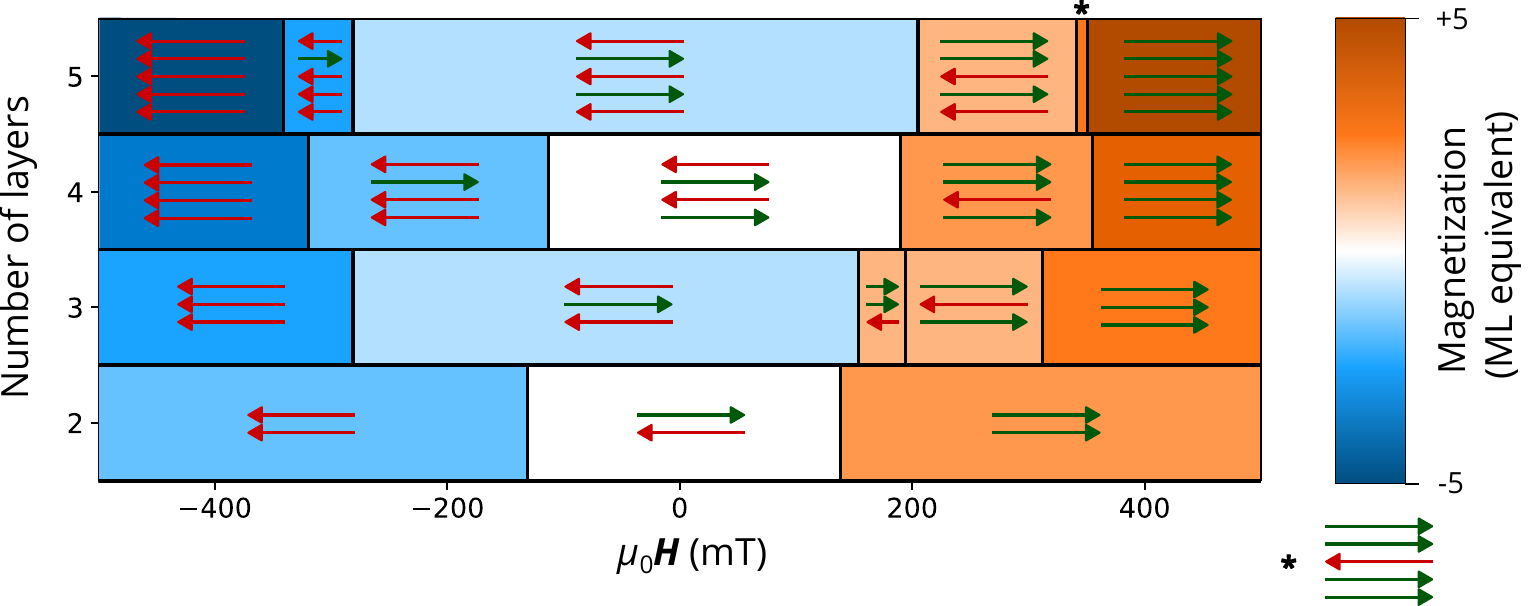}
    \caption{Spin-flip transition in few layers CrSBr. Arrows represent the magnetization of each layer while background color represent the total magnetization of the flake.}
    \label{Fig_spin_flip}
\end{figure}

Fig.\ref{Fig_spin_flip} represents the values of spin-flip transition reported in the literature. Bilayer spin-flip values where found in Wilson et al.\cite{Wilson2021Dec}, trilayer in Tabataba-Vakili et al.\cite{Tabataba-Vakili2024Jun} and  4 and 5-layer in Liu et al.\cite{Liu2024Jul}

Exact values for the spin flip of a given stack differ from study to study. 
For instance,  Wilson et al.\cite{Wilson2021Dec} reports flipping value between $130$ and $140~$mT for a bilayer, while Tabataba-Vakili et al.\cite{Tabataba-Vakili2024Jun} reports values up to $\sim 160~$mT.
These discrepancies could be explain by many factors, such as strain \cite{Cenker2022Mar}, lack of nucleation sites, or lateral exchange bias.

Particularly relevant for this work or the two arrangements of the trilayer ($\leftarrow$/$\rightarrow$/$\rightarrow$) and ($\rightarrow$/$\leftarrow$/$\rightarrow$) that were reported in \cite{Tabataba-Vakili2024Jun}.
These two stacking orders have the same total magnetization of $-1~M_{ML}$ and are therefore indistinguishable by a magnetometry measurement.
In our experiments however, both the contact angles and the presence or absence of an AF domain wall seemed to indicate that trilayer was systematically in the ($\rightarrow$/$\leftarrow$/$\rightarrow$) or ($\leftarrow$/$\rightarrow$/$\leftarrow$) state.

\newpage
\section{Field history}
\begin{figure}[h]
    \centering
    \includegraphics[width=\linewidth]{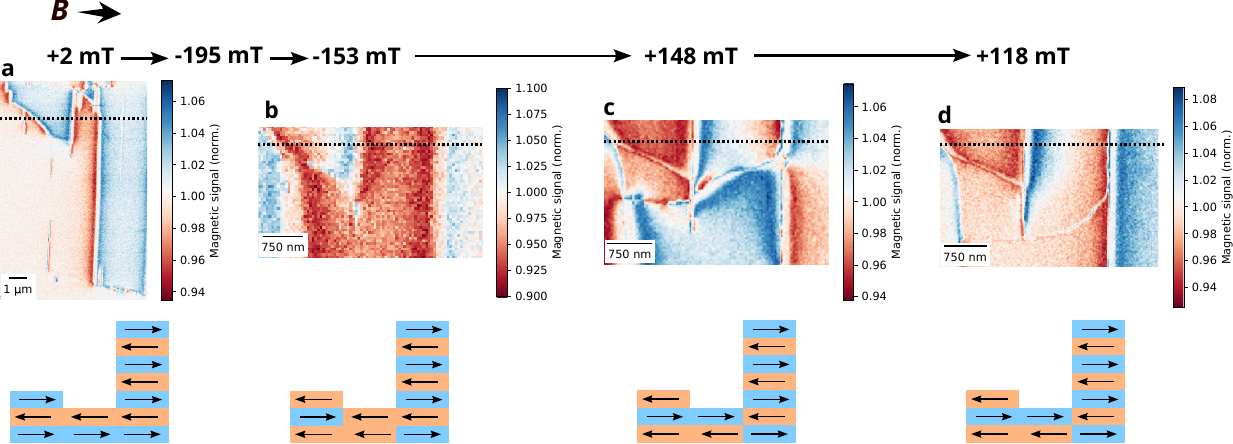}
    \caption{Dual-Iso-B magnetic imaging at various points during the magnetic field cycle used to create the AF domain wall. A cross-section illustrates the flake magnetic order for each field studied.}
    \label{Fig_Magnetic_history}
\end{figure}

Fig.\ref{Fig_Magnetic_history} details one of the field cycle used to create an AF domain wall in the bilayer.  The flake is first initialized so that each layer is uniformly magnetized. This is achieved either by field cooling or by applying a strong ($\sim 400~$mT) in plane positive field. Then a negative field of $-195~$mT is applied in order to flip the trilayer without flipping the 7-layer. Following this, we apply a positive field of $+148~$mT, near the bilayer spin-flip transition in order to assess the AF order by using the contact angle (see bellow) formed by the phase wall. If two different AF orders are observed, the field is then reduced to $+118~$mT in order to observe a domain wall.

In this scheme, it is not strictly necessary to switch from negative to positive field between step \textbf{b} and \textbf{c}, the domain wall should be present as soon as the bilayer is flipped back to AF order ($\approx -140~$mT). However for consistency, we always measure the domain wall with positive fields. Similarly, step \textbf{c} is technically facultative as the domain wall could be observed directly. However, since the stray field from the AF domain wall is much weaker than that of the AF/FM phase wall, it is easier to look for the deformations of the phase wall in order to assess whether or not a domain wall will be produced.

\newpage
\section{Contact angle and wetting analogy}
\begin{figure}[h]
    \centering
    \includegraphics[width=0.9\linewidth]{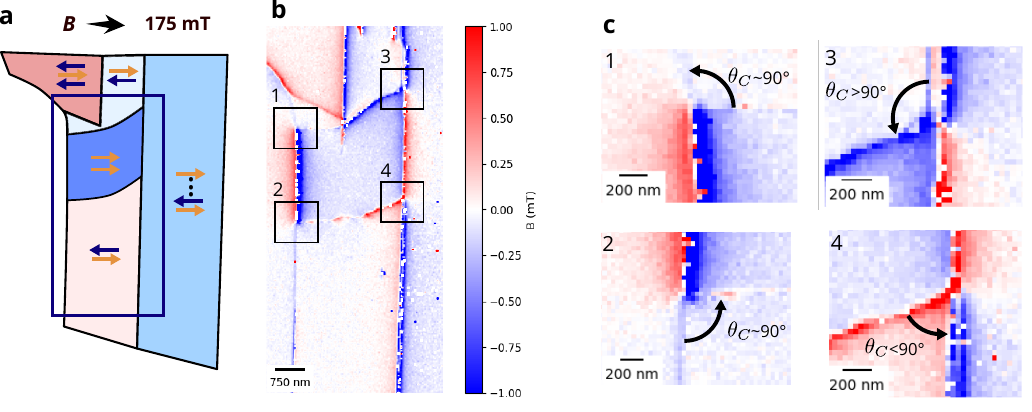}
    \caption{Contact angles at the bilayer$-$pinning layer interface. \textbf{a} Schematics of the flake's magnetization. \textbf{b} ODMR magnetic imaging of the flake. \textbf{c} ODMR magnetic zoom in on the four corners of the FM region.}
    \label{Fig_contact_angle}
\end{figure}

The distortions of the phase walls described in Fig. 2 of the main text can be seen as a form of ``contact angle'' in an analogy with the phenomenon of wetting in hydrostatics.
Indeed, we can assign a surface energy cost at each of the 3 interfaces (AF bilayer$-$pinning layer, FM bilayer$-$pinning layer and AF bilayer$-$FM bilayer), similarly to how wetting is defined by the surface energy between the liquid phase, the gas phase and the substrate.

It should be noted that the analogy between our observations and wetting is not perfect. For one, dipolar field (which is a long range interaction) plays a significant role in the shape of the domains, whereas surface tension is only defined by local interactions. Secondly, the volume of each phase (AF and FM) is not preserved in our experiments, unlike incompressible liquids.

Fig. \ref{Fig_contact_angle} shows ODMR maps of the sample in the state described in Fig. 2\,\textbf{a-d} of the main text, with a focus on the four corners of the FM phase. For the two left-most corners (1 and 2), a contact angle of $\sim 90 ^\circ$ can be observed. This configuration is expected in the absence of LEB as it minimizes both exchange energy\,\cite{Hubert2008a} and dipolar energy (since the phase wall runs parallel to the magnetization of the FM phase).

The two right-most corners however show clearly $\theta_C \neq 90 ^\circ$, as observed by the strong stray field on the phase wall. These two angles, and their opposite deviation from $\theta_C = 90^\circ$ , have been explained in the main text as a consequence of the Néel order in each AF regions and the LEB exerted by the pinning layer. Following with the wetting analogy, we can define angle \textbf{3} ($\theta_C > 90 ^\circ$) as ``hydrophobic'' because of the high surface energy between the AF phase and the pinning layer, and reversely angle \textbf{4} ($\theta_C < 90 ^\circ$) as ``hydrophilic''.

\newpage
\section{Additional contact angles and domain wall examples}
\label{sect:more examples}
\begin{figure}[htb!]
    \centering
    \includegraphics[width=0.6\linewidth]{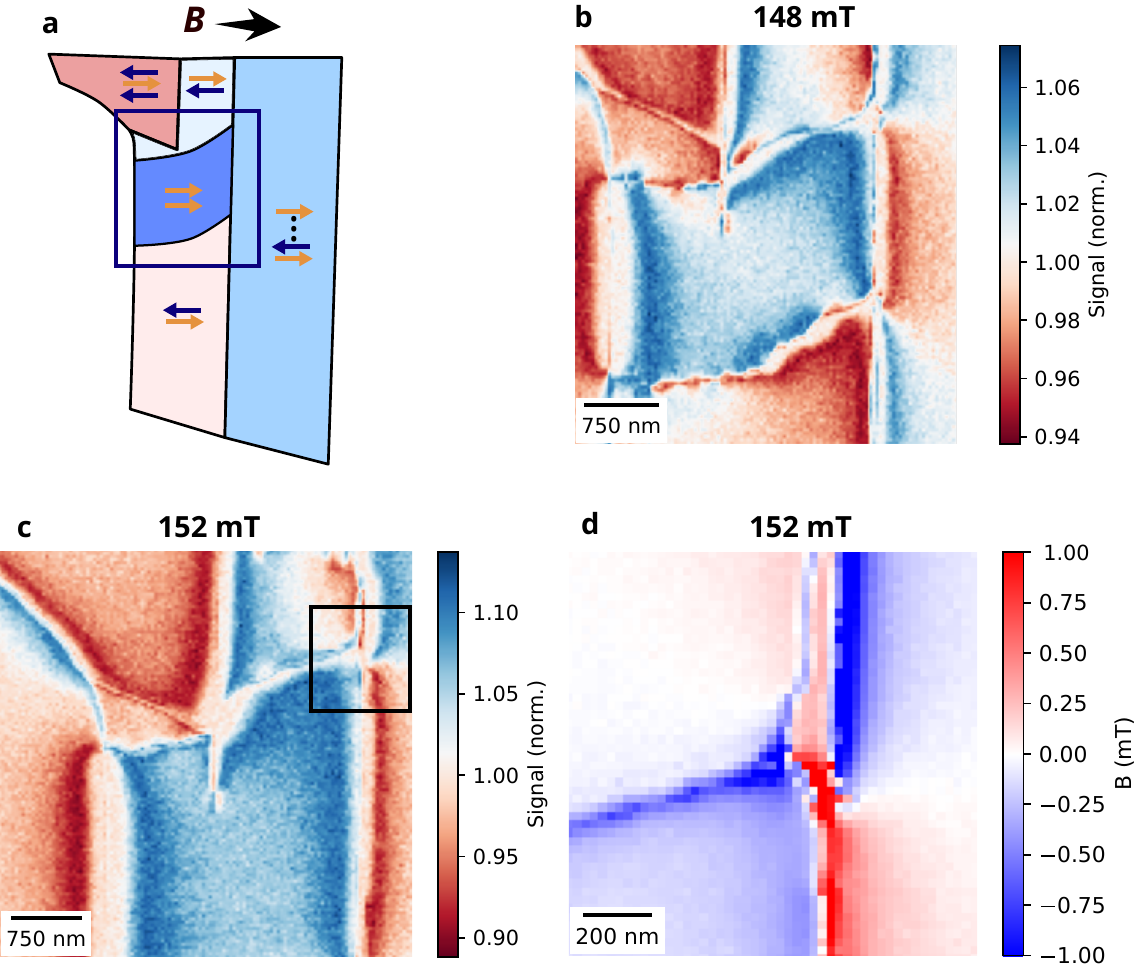}
    \caption{Example of $\theta_C>90^\circ$ (hydrophobic) contact angles. \textbf{a} Schematics of flake magnetization. \textbf{b} Dual-Iso-B magnetic imaging for an external in plane field of 148 mT. \textbf{c} Dual-Iso-B magnetic imaging for an in plane field of 152 mT after a different magnetic cycle. \textbf{d} ODMR magnetic imaging zoom-in from \textbf{c}.}
    \label{Fig_Wetting_S1_Hydrophobic}
\end{figure}

This section covers more examples of contact angles and AF domain walls.

Fig. \ref{Fig_Wetting_S1_Hydrophobic}\,\textbf{b} and \textbf{c} show magnetic images of the flake in the same condition as Fig.2,\textbf{b} of the main text after two distinct LEB cycles.
In both cases, we can observe a $\theta_C>90^\circ$ (hydrophobic) angle between the top AF domain, the FM region and the pinning layer.

Fig. \ref{Fig_Wetting_S1_Hydrophobic}\,\textbf{d} reveals a more complex internal structure to the hydrophobic contact angle.
Similarly to the micro-magnetics simulations (see section \ref{sect:micromagnetics}), we can observe that a small AF domain of N\'eel vector → (aligned with the pinning layer) forms along the edge of the pinning layer.
We can observe the AF domain wall propagating vertically along the 7-layer edge in both Fig. \ref{Fig_Wetting_S1_Hydrophobic}\,\textbf{c} and \textbf{d}.
As a result, there is a small internal hydrophilic contact angle inside the "macroscopic" hydrophobic wetting angle.
The same behavior can be seen in Fig.3\,\textbf{b} of the main text.

\begin{figure}[htb!]
    \centering
    \includegraphics[width=0.6\linewidth]{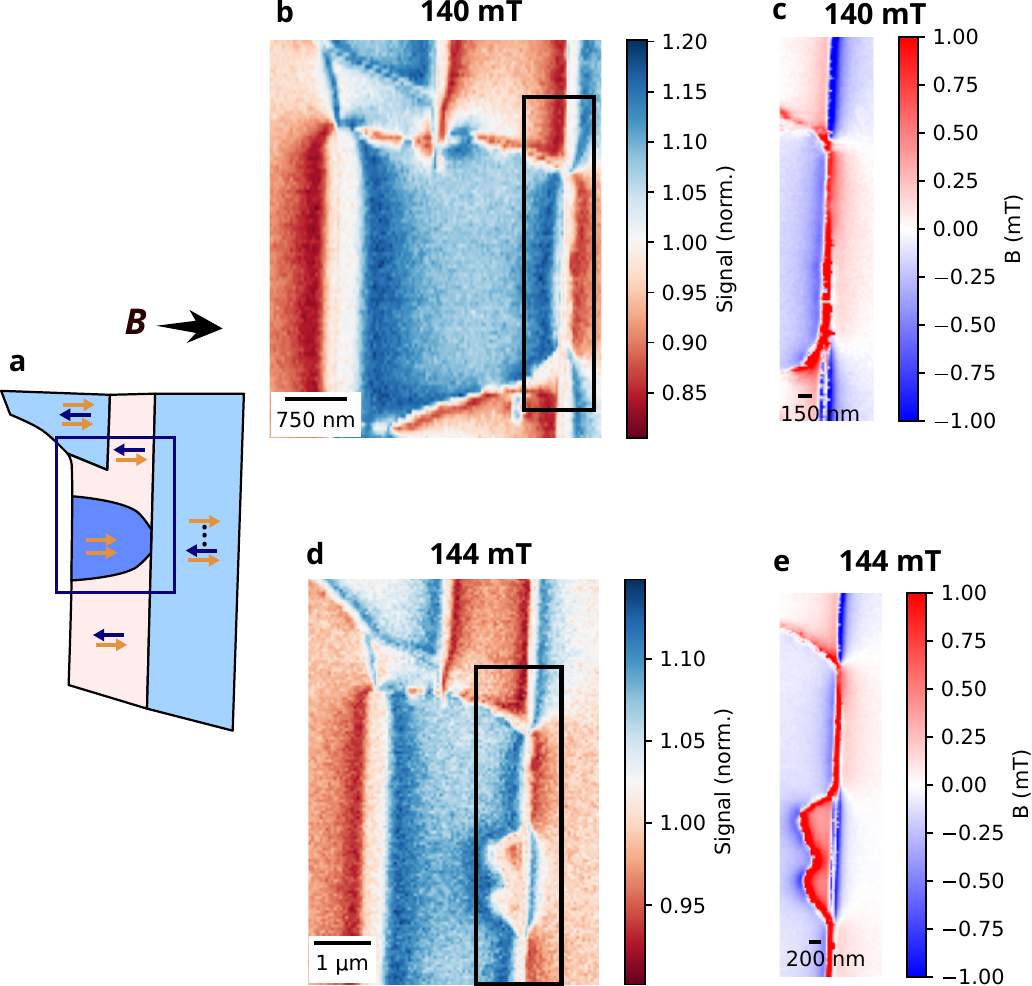}
    \caption{Example of $\theta_C<90^\circ$ (hydrophilic) contact angles on sample $S_1$. \textbf{a} Schematics of flake magnetization. \textbf{b(d)} Dual-Iso-B magnetic images for an external in-plane field of 140(144) mT after two different magnetic cycles. \textbf{c(e)} ODMR magnetic image zoom-in from \textbf{b(d)}.}
    \label{Fig_Wetting_S1_Hydrophilic}
\end{figure}

Fig.\ref{Fig_Wetting_S1_Hydrophilic}\,\textbf{b} and \textbf{d} show more examples of $\theta_C<90^\circ$ (hydrophilic) contact angles.
For these two series, an in-plane field of $\sim 300~$mT was first applied to create a homogeneous magnetization in all the layers.
As a result, the two contact angles visible in Fig.\ref{Fig_Wetting_S1_Hydrophilic}\,\textbf{b} and the 3 visible in Fig.\ref{Fig_Wetting_S1_Hydrophilic}\,\textbf{d} are all hydrophilic.
Unlike the hydrophobic case, the zoomed-in scans \textbf{c} and \textbf{e} do not reveal any internal structure for these hydrophilic angles.

\begin{figure}[htb!]
    \centering
    \includegraphics[width=0.7\linewidth]{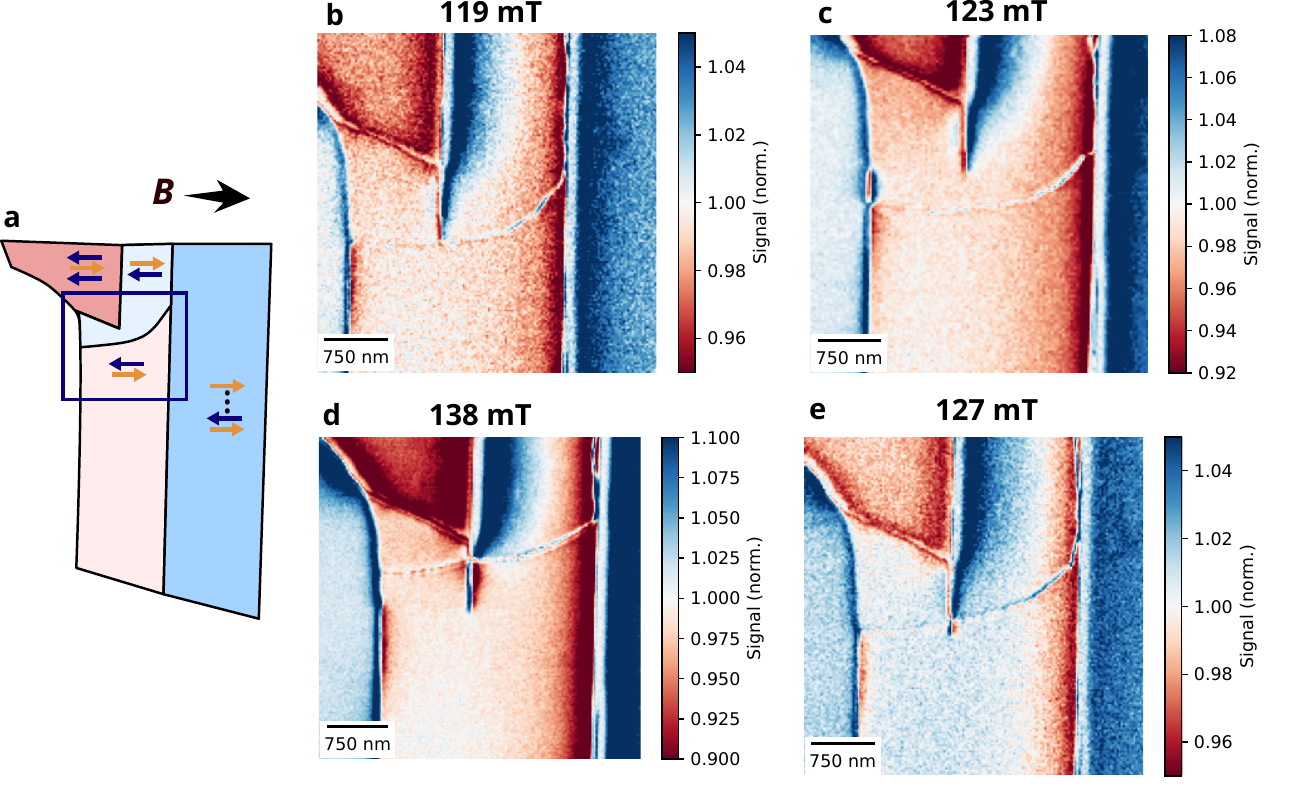}
    \caption{Example of domain walls on sample $S_1$. \textbf{a} Schematics of the flake and layer stacking order. \textbf{b,c,d,e} Dual-Iso-B images for 4 different LEB cycles. \textbf{d} corresponds to Fig. 4\,\textbf{a} from main text.}
    \label{Fig_S1_DW}
\end{figure}

Finally, Fig.\ref{Fig_S1_DW}\ shows the AF domain wall obtained after 4 different iterations of the LEB protocol.
The position of the domain wall is shifted from one iteration to the other, but the overall shape is consistent.
In particular, we can observe an angle minimizing the surface between the pinning layer and the AF2 domain (top).
We can also see a vertical domain wall propagating along the bilayer/7-layer edge, similar to what we described in Fig.\ref{Fig_Wetting_S1_Hydrophobic}.

\newpage
\section{Second sample}

\begin{figure}[h]
    \centering
    \includegraphics[width=0.7\linewidth]{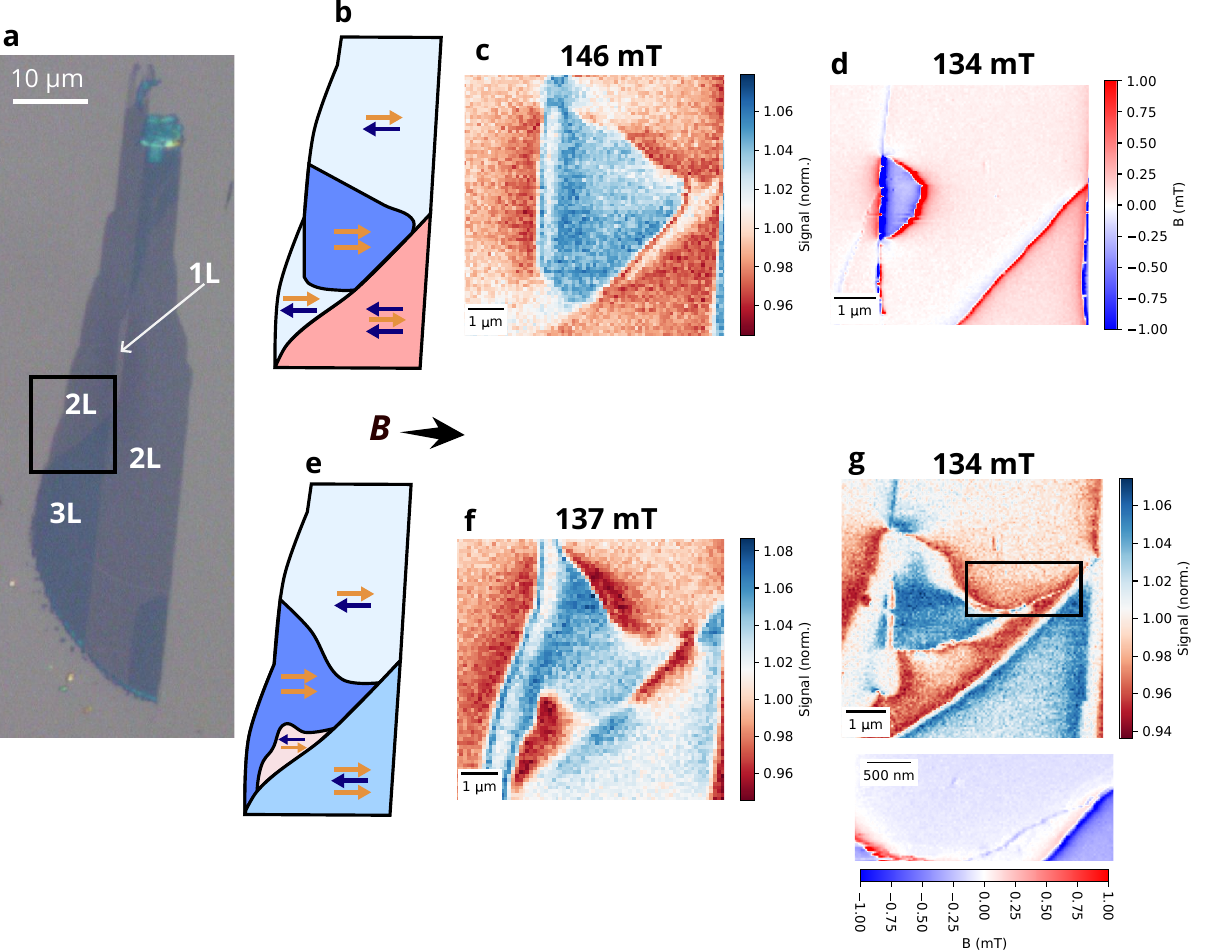}
    \caption{Lateral exchange bias on sample $S_2$. \textbf{a} Optical image of the flake with stacks of 1, 2 and 3 layers being labeled. \textbf{b} Schematics of the layers stacking order for the left bi- and trilayer (trilayer in the AF $M_S=-1~$ML state.) \textbf{c} Dual-Iso-B magnetic imaging (scanning window as a black rectangle in \textbf{a}) of the flake for an external in plane field of 146 mT and the trilayer in the AF $M_S=-1~$ML state. \textbf{d} ODMR magnetic imaging  of the flake for an external in plane field of 134 mT and the trilayer in the AF $M_S=-1~$ML state. \textbf{e} Same as \textbf{b} with the trilayer in the AF $M_S=+1~$ML state. \textbf{f} Dual-Iso-B magnetic image of the flake for an external in plane field of 137 mT and the trilayer in the AF $M_S=+1~$ML state. \textbf{g} (top) Dual-Iso-B magnetic imaging of the flake for an external in plane field of 134 mT and the trilayer in the AF $M_S=+1~$ML state. (bottom) ODMR magnetic image of the window shown on the the top image} 
    \label{Fig_Wetting_S2}
\end{figure}

The LEB protocol described in main text was applied to a second sample that we name $S_2$.
An optical image of this sample is shown in Fig.\,\ref{Fig_Wetting_S2}\,\textbf{a}.
The relevant stacks for this study are the bottom trilayer and the bilayer directly on top.
Similarly to the flake presented in main text, when applying an external field close to the bilayer spin-flip transition, the bilayer is split into a central FM region and a top and a bottom AF regions.

Fig.\,\ref{Fig_Wetting_S2}\,\textbf{b},\textbf{c} and \textbf{d} represent the case where the trilayer is in its $M_S=-1~$ML AF state.
Both contact angles visible in Fig.\,\ref{Fig_Wetting_S2}\,\textbf{c} are $<90^\circ$, meaning that both top and bottom regions have a N\'eel vector →.
This is confirmed in Fig.\,\ref{Fig_Wetting_S2}\,\textbf{d} by reducing the external field and observing that no domain wall appears.

On the other hand, Fig.\,\ref{Fig_Wetting_S2}\,\textbf{e},\textbf{f} and \textbf{g} represent the case where the trilayer is in its $M_S=+1~$ML AF state (flipped by applying a $+180~$mT in plane field).
There, the contact angle formed by the bottom AF region is $<90^\circ$, but the one formed by the top one is $>90^\circ$.
The top region therefore has a N\'eel vector ←, and the bottom region →.
This is again confirmed in Fig.\,\ref{Fig_Wetting_S2}\,\textbf{g} by reducing the field and observing the appearance of a domain wall between the two AF domains.

In both cases, the bottom AF region seems to nucleate from the trilayer boundary, which explains why it changed its order parameter when the trilayer was flipped.
The top domain AF however nucleates outside of the scanning window. 
We expect that it nucleates from some of the thicker flakes visible on the top of the optical image, as applying fields up to 300 mT was not enough to switch its AF order.

\bibliographystyle{naturemag}
\bibliography{CrSBr_bib}